\documentclass[aps,pra,superscriptaddress,nofootinbib,showpacs,longbibliography,twocolumn,10pt]{revtex4-1}

\usepackage[utf8]{inputenc}
\usepackage[english]{babel}
\usepackage{times}
\usepackage{color}
\usepackage{graphicx}
\usepackage{import}
\usepackage{amsmath}
\usepackage{braket}
\usepackage[caption=false]{subfig}

\usepackage{booktabs}

\AtBeginDocument{
\heavyrulewidth=.08em
\lightrulewidth=.05em
\cmidrulewidth=.03em
\belowrulesep=.65ex
\belowbottomsep=0pt
\aboverulesep=.4ex
\abovetopsep=0pt
\cmidrulesep=\doublerulesep{}
\cmidrulekern=.5em
\defaultaddspace=.5em
}

\usepackage{tikz}
\usepackage{tikzscale}
\usetikzlibrary{backgrounds,fit,decorations.pathreplacing,calc,shapes}

\newcommand\encircle[1]{%
  \tikz[baseline= (X.base)]
    \node(X) [draw, shape=circle, inner sep=-0.5pt] {\tiny\strut#1};
}

\begin{document}
\title{Lattice Surgery Translation for Quantum Computation}
\author{Daniel Herr}
\email{daniel.herr@riken.jp}
\affiliation{Quantum Condensed Matter Research Group, CEMS, RIKEN, Wako-shi 351-0198, Japan}
\affiliation{Computational Physics, ETH Zurich, 8093 Zurich, Switzerland}
\author{Franco Nori}
\affiliation{Quantum Condensed Matter Research Group, CEMS, RIKEN, Wako-shi 351-0198, Japan}
\affiliation{Department of Physics, University of Michigan, Ann Arbor, MI 48109-1040, USA}
\author{Simon J. Devitt}
\affiliation{Quantum Condensed Matter Research Group, CEMS, RIKEN, Wako-shi 351-0198, Japan}
\affiliation{Superconducting Quantum Simulation Research Group, CEMS, RIKEN, Wako-shi 351-0198, Japan}
\begin{abstract}
  In this paper we outline a method for a compiler to translate any non fault tolerant quantum circuit to the geometric representation of the lattice surgery error-correcting code using inherent merge and split operations. Since the efficiency of state distillation procedures has not yet been investigated in the lattice surgery model, their translation is given as an example using the proposed method. The resource requirements seem comparable or better to the defect-based state distillation process, but modularity and eventual implementability allow the lattice surgery model to be an interesting alternative to braiding.
\end{abstract}


\maketitle
\section{Introduction}
  Performing any type of quantum computation is a delicate undertaking. Quantum systems are easily perturbed due to environmental influences and/or imperfect control and thus lead to unwanted changes in the physical qubits, which will in turn lead to computational errors.

  Quantum Error-Correction (QEC) has become an extremely well developed component of quantum information science and has shown how arbitrary quantum algorithms can be realized provided physical error rates of the hardware are below a certain threshold.  While there are multiple ways to implement QEC, with numerous codes and fault-tolerant protocols available~\cite{qec_book,LB13,G09}, quantum engineers need to consider hardware constraints when designing practical large-scale hardware that will not require a physically unreasonable number of physical qubits or computational time to realize an error-corrected algorithm~\cite{DSMN13,AMK15,LHMB15}.

  The toric code~\cite{Kitaev2003} was the first topological quantum error-correction code discovered that had the potential for a realistic hardware implementation while also having comparatively good performance (in terms of the fault-tolerant threshold). In this code, physical spins are arranged on a two dimensional lattice with periodic boundary conditions. A set of commuting quantum operators, called stabilizers, is chosen and measured continuously.  The stabilizers are defined locally over a group of four neighbouring spins and the continuous measurement of the eigenvalue of these stabilizers enables the detection and correction of errors on physical spins. For a $N$-qubit toric code, there are $(N-2)$ linearly independent stabilizers, hence there are two degrees of freedom that can be used to encode information. Thus, the toric code can be used to encode two logical qubits.
\subsection{Toric Code}
  The structure of the toric code (most notably the periodic boundary conditions) makes this code difficult to implement in realistic hardware. Luckily, this code can be generalized to other variants. The most common is the surface code~\cite{DKLP02,FSG08,Fowler2012}, which is still defined on a 2D lattice of qubits, but does not require periodic boundary conditions. Additionally, the number of degrees of freedom (encoded qubits) can be greater than two, with a sufficiently large array of physical qubits, by voluntarily switching off stabilizer measurements. These are termed defects and any computation can be performed by braiding them around each other~\cite{Fowler2012,RHG_NJP}.
\subsection{Surface Code}
  The surface code itself, and its performance, has been extensively studied~\cite{BD+07,WFSH09,BAO12,S14,F15}, and due to both its performance and comparative ease of implementation for physical hardware has recently become the code of choice for most hardware models under experimental development~\cite{JMFMKLY10,D09,MRRBMD14,NTDS13,LWFMDWH15,HPHHFRSH15}. There has also been significant work related to the classical compilation and control for a quantum computer operating under this model~\cite{DFTMN10,GP10,PF13,D14,F15+}. Compiling fault-tolerant quantum algorithms for this model essentially consists of generating a 3-dimensional geometric description of a topological circuit which has a certain space/time volume~\cite{FD12}. Resource optimization requires the compaction of this structure using rules that manipulate and reduce this space/time volume (and consequently the number of qubits/time required for computation) without changing the functionality of the topological circuit~\cite{PF13}. However, this optimization problem has so far proven to be difficult~\cite{PDF16} and the resource cost is still unclear for fully compiled and optimized quantum circuit.

  \begin{table*}
  \setlength{\tabcolsep}{8pt}
  \renewcommand{\arraystretch}{2}
    \begin{tabular}{p{4.5cm} p{0.1cm} p{2.5cm} p{3.2cm} p{4cm}}
      \toprule
      && {\bf Toric Code} & {\bf Surface Code} & {\bf Planar Code\newline (Lattice Surgery)}\\
      \midrule
      {\bf Encoding qubits} && Loops around torus & Defects & Patch of planar code with open boundary conditions \\
      {\bf Number of possible qubits} && 2 & $\infty$ & 1 per patch \\
      {\bf Performing computation} && Memory only & Braided logic& Merge and split operations\newline between isolated planar codes \\
      {\bf Compiler to geomentric\newline representation} && Not applicable & Was devised in\newline Ref.~\cite{PPND15,PDF16}& This work \\
      {\bf Optimization} && Not applicable & Unsolved & Conceptually simpler \\
      {\bf Overall space/time requirements for $\ket{Y}$-state distillation} && Not applicable & $140\, d^3$\, Ref.~\cite{FD12} & $280 \, d^3$ \\
      {\bf Overall space/time requirements for $\ket{A}$-state distillation} && Not applicable & $1500\, d^3$\, Ref.~\cite{FD12} & $1080 \, d^3$ \\
      {\bf Overall space/time requirements Bravyi-Haah $(3k + 8)$-to-$k$ state distillation for $k=4$} & & Not applicable & $4688\, d^3$\, Ref.~\cite{bravyi_resource} & $2268 \, d^3$ \\
      \bottomrule
    \end{tabular}
  \caption{\label{fig:err_overview}Overview between different topological error-correction techniques. Here, $d$ denotes the error-correcting code distance. The space-time requirements for lattice surgery are calculated at the end of this paper.}
  \end{table*}

\subsection{Planar Code}
  A second approach to achieve universal computation with the surface code does not rely on topological braiding. This type of toric code is the planar code~\cite{BK01}. It allows for the encoding of a single qubit of information without periodic boundary conditions; hence the computer now becomes an array of isolated 2D patches of surface code, each representing a logical qubit. The toric code and its variants allow for a transverse application of the logical CNOT gate (a transversal CNOT gate is where corresponding physical qubits in two logical blocks are interacted via a physical CNOT gate), but a transversal logic gate eliminates the 2D nearest-neighbor geometry that is extremely desirable for hardware implementation. To mitigate this problem a technique, called \textit{lattice surgery}~\cite{Horsman2012}, was developed, that re-introduces only 2D nearest-neighbor interactions to achieve a logical CNOT gate between two encoded qubits.
  These CNOTs are achieved by turning on (off) syndrome measurements on the boundary between adjacent qubits. Such operations are called \textit{merges (splits)}. A computation in this geometric structure has to perform computation using merges and splits, which can emulate gates such as CNOTs. Combining this with state-injection~\cite{BK05+} allows for the realization of a universal set of quantum gates~\cite{Horsman2012}.

  Circuit compilers have been developed that translate higher-level circuits into the appropriate geometric forms for braid-based topological computation~\cite{PPND15,PDF16}. Preliminary steps have been made to both benchmark and optimize physical resources using this model~\cite{DSMN13}, but the question remains whether, ultimately, a lattice-surgery-based approach to computation will be more resource efficient. As with braiding-based computation, we first need a generalized set of protocols to compile an appropriate fault-tolerant circuit into a form appropriate for lattice surgery, design optimization protocols for this form and ultimately compare physical resources to braiding-based approaches. In this paper we take the first step and provide a method that will convert an appropriately designed high-level circuit into a physical layout and scheduling pattern for implementation in a lattice-surgery-based topological quantum computer.

\subsection{ICM Representation}
  For this we start with the ICM representation~\cite{PPND15}, which is a formulation used to compile arbitrary circuits using braid-based logic, and is divided into three distinct parts: {\bf I}nitializations, {\bf C}NOTs and {\bf M}easurements. This description operates at the logical level of the computation and is designed to be compatible with all Calderbank-Shor-Steane-based error-correction codes~\cite{CSS,Nielsen2011}.  First, all qubits are initialized in one of four distinct states. Secondly, CNOTs are applied to these qubits to perform entanglement operations; then in the last step the qubits are measured.

  This formalism incorporates not only the higher-level decompositions to convert a quantum algorithm into an appropriate Clifford $+T$ gate library, compatible with fault-tolerant error-correction, but also includes ancillary protocols such as state distillation~\cite{Statedistill_meier,Statedistill_BravyiHaah,Statedistill_Cody,BK05+}, which are needed for an operational computer. 
  Our formalism utilizes an inverse model of ICM~\cite{RotatedICM}, where qubit initializations are restricted to two states, $\ket{0}$ and $\ket{+}= \frac{1}{\sqrt{2}}\left(\ket{0}+\ket{1}\right)$, and measurements are performed in a rotated basis to achieve universality.
  The entangled states before their measurements in the inverted ICM representation is similar to graphs states and their creation using parity checks has been described in~\cite{fusion_operators,MB_implementation}. 
  By working in this inverted ICM representation, we can use the natural operations exhibited by the lattice surgery protocol to realize a universal set of logic operations and to layout and schedule an arbitrary quantum circuit for implementation on an actual error-corrected machine.

  The inverted ICM formalism follows a similar structure to measurement-based quantum computation~\cite{MBQC_oneway,MBQC_briegel}; however, our approach works at the fault-tolerant error corrected level. This formalism prepares an effective encoded graph state which is algorithmically-specific given the original circuit specification. Computation then proceeds via encoded rotated-basis measurements on each qubit of this encoded graph. Our method prepares such an entangled state in a completely fault-tolerant manner during the initialization and CNOT steps and maintains the physical 2D nearest-neighbor restrictions of the underlying hardware.

  The work presented in this paper is akin to the already existing compiler for the braided error-correction scheme~\cite{PDF16}. Using this we will discuss its resource requirements on three exemplary state-distillation algorithms that we have attempted to optimize manually. As the bulk of operations in a fault-tolerant quantum computer is related to ancillary protocols, such as state distillation, numerous techniques have recently been investigated~\cite{Ogorman,Campbelldistill}. To illustrate the translation, this is done explicitly, and its complexity is compared to the braiding method.

  Nevertheless, the three examples of state distillation circuits in this work were implemented naively for illustrative purposes and did not take advantage of recently proposed methods of optimization~\cite{Campbelldistill,CO16}. Additionally, there have been several papers that have further optimized the layout of encoded information using the planar code~\cite{NSM16,DIP16}. These proposed techniques are compatible with lattice surgery and hence the results presented in this work can be improved upon such that the resource requirements decrease even further. However, this analysis together with further scheduling optimizations are left for future work.

\subsection{Outline of the paper}
  We will now give a short outline on the structure of the paper and describe what has been done throughout the following sections. In section~\ref{sec:QuantumComputing} and section~\ref{sec:LS} we review previous literature and adapt their findings such that they can be used in our translation. Afterwards, in section~\ref{sec:ICMRep} we describe the format in which a quantum algorithm has to be represented in order to use our translation. In section~\ref{sec:IC} we outline the first part of our mapping to the fault-tolerant lattice surgery model. This description builds on the concepts presented in the previous sections. Afterwards we describe how to implement an algorithmically-specific entangled state, which in many error correcting frameworks is described by the stabilizer matrix formulation. In section~\ref{sec:stabilmatrix} we describe how our approach can be translated to this formalism, which allows easier comparison of the entangled states.
  To finish the translation to lattice surgery, we describe the procedure of measurement in section~\ref{sec:measurement}. This concludes the translation and three example (state distillation) circuits are investigated under the devised method in section~\ref{sec:example}. Section~\ref{sec:conclusion} summarizes our results.

\section{\label{sec:QuantumComputing} Quantum Computing}
 In general, error-correction schemes are agnostic of the underlying hardware. While codes are often constructed with hardware constraints in mind, any type of qubit in any type of appropriate hardware model can be used. In recent years, many advances in quantum hardware~\cite{laser_err_cor,superconducting_err_cor} have been made such that lower error rates in physical qubits were archived. Thus the threshold for surface code error-correction was surpassed and the implementation of surface code corrected qubits is now, in principle, possible (there is still significant work that is needed to scale systems and maintain low error rates). This makes the current investigations for this method of error-correction an important task. It involves providing a classical framework at the hardware level~\cite{physical_impl} to develop classical algorithms to track errors and correct them if necessary~\cite{SC_ClassProc,D14,F15+} and to optimize qubit and time resources for a compiled, error-corrected algorithm~\cite{PF13,FD12}.

 A universal quantum computer requires a certain, discrete, number of gates that can be used to construct any arbitrary unitary operation.  While there are many universal gate sets, there are restrictions imposed by a QEC code such that some sets are preferred over others. Most importantly, each element of a universal set needs a fault-tolerant implementation on an appropriately chosen quantum code. Arguably, the most common universal set (and the one that is used with the surface code) is the Clifford$+T$ set~\cite{cliff_T2,cliff_T,CliffordT,CliffordT2}, which is generated by the gates~\cite{Nielsen2011}:
  \begin{gather*}
    H = \frac{1}{\sqrt{2}} \begin{bmatrix}1 &1\\1&-1\end{bmatrix} \qquad
    T = \begin{bmatrix}1 &0\\0&e^{i\frac{\pi}{4}} \end{bmatrix}\\
    \text{CNOT} = \begin{bmatrix} 1&0&0&0\\
                                  0&1&0&0\\
                                  0&0&0&1\\
                                  0&0&1&0\end{bmatrix}.
  \end{gather*}
  While not strictly required, this generating set is generally augmented with the $P=T^2$ gate (which combined with the CNOT and Hadamard gate generates the Clifford group), as there are more efficient implementations of this gate compared to using two $T$-gates in sequence~\cite{BK05+}.

  It is common to write quantum algorithms in a circuit where each element is one of these operations. This is a good representation on a conceptual level, but needs to be translated to lower hardware. Such a feat is done by a quantum compiler. Among recent proposals~\cite{haener2016,PPND15,RotatedICM}, one proposed design stack consists of the following steps:
  \begin{enumerate}
    \item Algorithm: high-level functions like Quantum Fourier Transform, which consist of many applications of gates.
    \item Non-fault-tolerant circuit with general gate set: This representation is akin to the description of quantum circuits in textbooks.
    \item Fault-tolerant circuit with a reduced gate set: using a subset of gates and a defined structure, which can be implemented on the error-correcting code that will be used. One of these circuit classes is the ICM representation~\cite{PPND15}.
    \item Geometry: This is where the fault-tolerant circuits are translated into a geometric representation for braided topological logic~\cite{PDF16}. In this work we address this step for the lattice surgery approach using planar codes.
    \item Mapping: Measurement operations of syndrome qubits are switched off and on in order to perform the geometric algorithms~\cite{Braiding_compiler}.
    \item Hardware: Individual hardware instructions, e.g.\ laser pulses, that manipulate the state of single physical qubits.
  \end{enumerate}
  Our method translates from level 3 to level 4, where a general quantum circuit is transformed to a lattice surgery error-corrected algorithm. The description of the algorithm presented in this paper is still agnostic to its underlying hardware. For defect-based surface codes these algorithms already exist and were proposed in~\cite{PPND15}. However, the braided geometric representation has proven to be difficult to optimize~\cite{PF13} and has not been studied in depth.

  The method proposed in this paper, albeit a complex optimization problem, appears to be more feasible to optimize than the braided version, because it is relatively close to two problems, where one is the traveling salesman problem and the other the sliding puzzle problem~\cite{H86}.

\section{\label{sec:LS} Computation using Lattice Surgery}
\begin{figure}
    \centering
    \includegraphics[width=0.5\columnwidth]{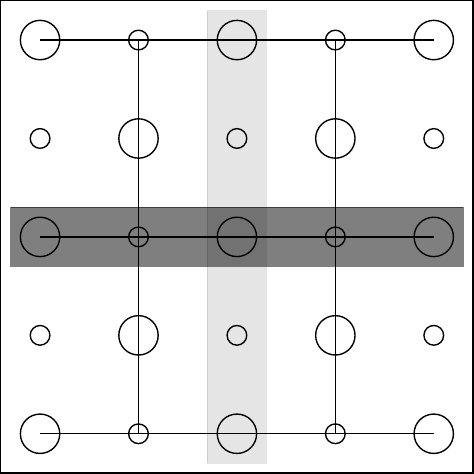}
    \caption{\label{fig:LSoperators} A depiction of a patch of error-corrected distance-3 surface code encoding one logical qubit. Here, the syndrome qubits are represented using small circles and data qubits are drawn in big circles. The application of a logical $X$-operator is given by performing a physical $\sigma_x$ operator on all data qubits in the light gray box. The logical $Z$-operator is performed by applying $\sigma_z$ on all data qubits inside the dark horizontal box.}
  \end{figure}

  We will now review the basic concepts of lattice surgery~\cite{Horsman2012} and in subsection~\ref{chap:merge_split} describe a method for entanglement creation used throughout our translation.
  In lattice surgery each qubit is encoded in one patch of error-correcting surface code with open boundary conditions (commonly referred to as a planar code). Logical $X$- and $Z$-operators are defined as chains of physical operators that span the whole patch of code from either left to right (logical $Z$) or top to bottom (logical $X$) [Figure~\ref{fig:LSoperators}]. Computation can be performed by allowing interaction between different patches of code.

  To interact two logical qubits, a line of data and syndrome qubits is added on the boundary between two patches of code and stabilizers between them are either turned on or off.
  Switching on/off these stabilizers, merging/splitting individual logical qubits, act as parity checks on the basis states of the logical qubits that partake in the operation. Using these operations natively instead of emulating the true effect of CNOT operators gives rise to more efficient computation. The four new operations are smooth/rough merges and smooth/rough splits. In the following, the effect of these operations is described, a more elaborate explanation can be found in the original paper~\cite{Horsman2012}.
\begin{figure}
  \subfloat[Rough merge\label{fig:rough}]{
  \includegraphics[width=0.6\columnwidth]{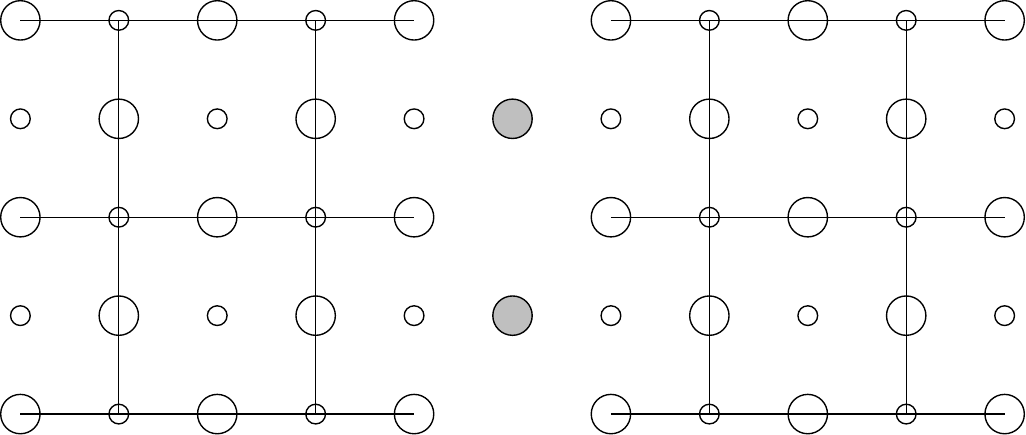}
  }
  \hfill
  \subfloat[Smooth merge\label{fig:smooth}]{
  \includegraphics[width=0.25\columnwidth]{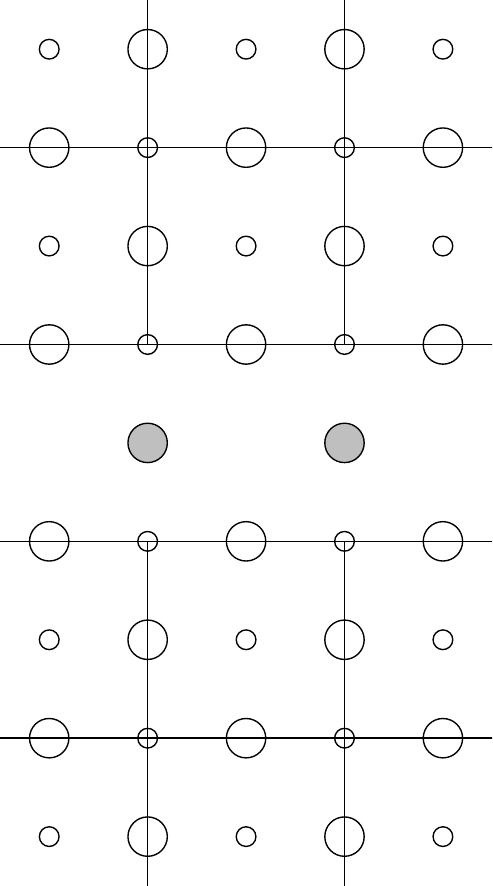}
  }
  \caption{\label{fig:smoothrough} Examples of the merge/split operation in lattice surgery codes. In this case, two stabilizer measurements, represented by the two gray circles are turned on (merge) or off (split). During the merge process these intermediate gray qubits need to be prepared in either $\ket{0}$ for a rough merge or $\ket{+}$ for a smooth merge.}
\end{figure}

\subsection{Merge operations}
  Figure~\ref{fig:smoothrough} shows the difference between a smooth and a rough merge in the lattice surgery model. One can see that this difference comes from stabilizer measurements that have to be turned on as a mediator between the different patches of surface code. For a smooth merge these are $Z$-stabilizers, whereas for a rough merge $X$ stabilizers are used.
  As an example, we describe the rough merge process. First, the intermediate (physical) data qubits need to be prepared in the $\ket{0}$ state, then the measurements of the $X$-stabilizers along the edge between the two surfaces are turned on. These measurements will later be needed to precisely determine the state and whether a correction operator needs to be applied.

  Mathematically, if the initial states are given by ${\ket{\psi}= \alpha \ket{0} + \beta\ket{1}}$ and ${\ket{\phi} = \alpha' \ket{0} + \beta' \ket{1}}$, the post-merge state for a rough merge evaluates to
  \begin{align*}
    \ket{\psi} \encircle{M}_r \ket{\phi} &= \alpha \ket{\phi} + {(-1)}^M \beta \ket{\overline{\phi}} \\
    & = \alpha' \ket{\psi} + {(-1)}^M \beta' \ket{\overline{\psi}}
  \end{align*}
  where $\ket{\overline{\phi}} = \sigma_x \ket{\phi}$; and $M \in \{0,1\}$ is the aforementioned measurement result of the intermediate stabilizers. The effect of this operation is a parity measurement on both states, which decreases the number of possible degrees of freedom by one. For the smooth merge, one has to consider a basis transformation of the pre-merge states to $\ket{\psi} = a\ket{+} + b\ket{-}$ and $\ket{\phi} = a'\ket{+} + b' \ket{-}$. The post-merge state now evaluates to
  \begin{align*}
    \ket{\psi} \encircle{M}_s \ket{\phi} &= a \ket{\phi} + {(-1)}^M b\ket{\overline{\phi}} \\
    & = a' \ket{\psi} + {(-1)}^M b' \ket{\overline{\psi}}.
  \end{align*}
  Here the state $\ket{\overline{\phi}} = \sigma_z \ket{\phi}$ is the negation in the $\pm$ basis.

  \subsection{Split operations}
  During a split operation a single logical qubit will be split into two. Again the boundary between the two newly created logical qubits will be used to discriminate between a smooth and a rough split. The individual qubits of this border are measured out and the two remaining surfaces are then stabilized individually.

  For a smooth split we can note that the removal of the qubits on the boundary will not change the outcome of any of the joint $Z$-operators. Thus both states are in the same superposition of eigenstates of the $Z$-operator. Mathematically, the smooth split will give
  \begin{equation}
    \alpha \ket{0} + \beta \ket{1} \rightarrow_s \alpha \ket{00} + \beta \ket{11}.
  \end{equation}
  For the rough split one will get
  \begin{equation}
    a \ket{+} + b \ket{-} \rightarrow_r a \ket{++} + b \ket{--}.
    \label{eq:roughsplit}
  \end{equation}
  Performing a basis transformation on this last equation will give us the effect of a rough split on an arbitrary state in the $Z$-basis
  \begin{equation}
    \alpha \ket{0} + \beta \ket{1} \rightarrow_r \alpha \left(\ket{00} +\ket{11}\right)/\sqrt{2} + \beta \left( \ket{01} + \ket{10}\right)/\sqrt{2}.
    \label{eq:smoothsplit}
  \end{equation}
  These split operations enable the creation of entanglement among the encoded qubits.
  One should note that after performing a split or merge one needs $d$ rounds of error-correction cycles to compensate for faulty physical measurements in the computer. However, this might be reduced by making use of the fact that the errors are likely concentrated along the boundary of the split and merge, and this is currently under investigation.

  \subsection{Multi-target CNOT}
    \begin{figure}
    \centering
    \includegraphics[width=\columnwidth]{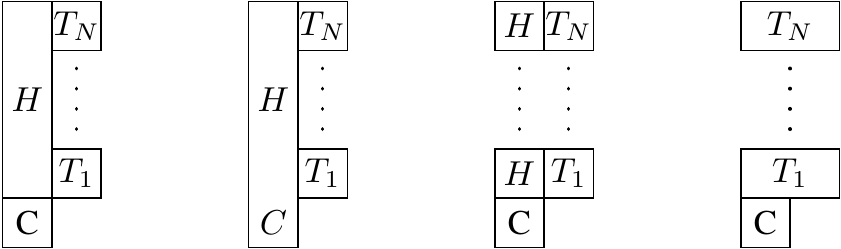}
    \caption{\label{fig:multi_qubit_cnot} This figure illustrates the application of multi-target CNOTs in lattice surgery. The patch denoted by C corresponds to the control qubit and patches with $T_i$ denote the $i\text{th}$ target. Furthermore, patches denoted by $H$ are ancillary patches. In a first step, a vertical patch composed of all ancillas is initialized to $\ket{+}$ and using a smooth merge the control qubit is merged to this patch. Afterwards, this large section is split into individual patches and in the final step each ancilla is merged to one target. This operation results in the same state as a multi-target CNOT.}
  \end{figure}
  We now turn our attention to the implementation of logical operations. In the following, we introduce an implementation of fanouts or multi-target CNOT gates. These operations will form the core of our compiler.
  In order to properly perform a multi-target CNOT in lattice surgery, we expand the original description of Ref.~\cite{Horsman2012} on how to perform CNOTs. Here, one needs an additional encoded ancilla qubit for each target qubit. In the following derivation, we consider a general control qubit given by $\ket{\psi} = \alpha \ket{0} + \beta \ket{1}$. As depicted in Figure~\ref{fig:multi_qubit_cnot}, we prepare all encoded ancilla qubits in one single $\ket{+}$ state and perform a smooth merge between the ancilla qubit and the control qubit. After $d$ rounds of error-correction this combined qubit is split smoothly into $N+1$ qubits, where $N$ is the number of target qubits. The resulting state is given by ${\alpha \ket{0\cdots0} + \beta \ket{1\cdots1}}$. Now for each of the $N$ target qubits a rough merge with one of the entangled qubits from the smooth split is performed. This results in the state:
  \begin{align*}
    \ket{\psi} &= \alpha \ket{0} \otimes \left[\left(\ket{0} \encircle{M}_r \ket{T_1} \right) \cdots \left(\ket{0} \encircle{M}_r \ket{T_N}\right)\right] + \\
    &+ \beta \ket{1} \otimes \left[\left(\ket{1} \encircle{M}_r \ket{T_1} \right) \cdots \left(\ket{1} \encircle{M}_r \ket{T_N}\right)\right]
  \end{align*}
  The resulting state from the rough merge is given by
  \begin{equation*}
    \ket{\psi} = \alpha \ket{0} \otimes \ket{T_1}\otimes \cdots\otimes \ket{T_N} + \beta \ket{1} \otimes \ket{\overline{T_1}}\otimes \cdots \otimes\ket{\overline{T_N}}
  \end{equation*}
  which is exactly the effect of a multi-target CNOT.\@ A more resource-friendly implementation is discussed in the next section.

\subsection{\label{chap:merge_split} Split and Merge circuit}\label{chap:CNOT}
\begin{figure}
    \centering
    \includegraphics[width=0.7\columnwidth]{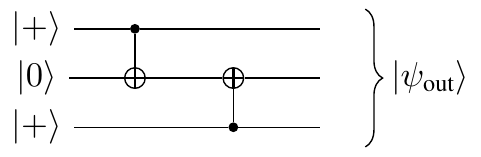}
    \caption{\label{fig:CNOTex} This circuit will be used as an example to see how CNOTs can be implemented in lattice surgery with less overhead. Furthermore, this can be extended to multi-target CNOTs.}
  \end{figure}

\begin{figure}
    \includegraphics[width=\columnwidth]{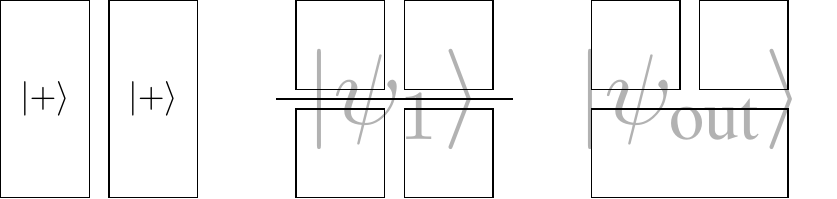}
    \caption{\label{fig:CNOTex2} Implementation of the circuit in Figure~\ref{fig:CNOTex} using the shorthand implementation of the multi-target CNOT operation (in this case only 1 target) for qubits. Three steps are needed: First, an initialization of two patches to $\ket{+}$; these correspond to the control qubits of each CNOT. Then, two smooth splits, represented by the horizontal line in the second frame, creating two two-qubit entangled states. In the end, a rough merge between two qubits connects two patches, such that the final state of the 3 remaining patches is the same as for the circuit in Figure~\ref{fig:CNOTex}\vspace{5mm}}
  \end{figure}

  The multi-target CNOT implementation introduced in the previous section works well but requires many ancillary patches. In this section we devise a shorthand implementation for entanglement creation by providing an example circuit in Figure~\ref{fig:CNOTex}. The drawback of this implementation is, however, that it is restricted to only $\ket{0}$ and $\ket{+}$ input states. The output state of this circuit is given by
  \begin{equation*}
    \ket{\psi_\text{out}} = \frac{1}{2} \left(\ket{000}+\ket{110}+\ket{011}+\ket{101}\right).
  \end{equation*}
  It can be seen in the following that the application of two splits and one merge has the same effect as this circuit. A graphical representation of this is given in Figure~\ref{fig:CNOTex2}. Two encoded qubits are both initialized to $\ket{+}$. Since these are encoded in rectangular patches of surface code, our schematic representation of this will be boxes which can be merged and split. Both of these qubits will now be split using a smooth split. This will lead to an intermediate state given by:
  \begin{equation}
  \ket{\psi_1} = \frac{1}{2} \left(\ket{00} + \ket{11}\right) \otimes \left(\ket{00} + \ket{11}\right)
  \end{equation}
  Using a rough merge between two of these patches, one will also obtain $\ket{\psi_\text{out}}$.

  The measurement during the split operations corresponds to measuring the operator $X_L X_L$ on the two qubits. Here, the measurement outcome is completely determined, and one will always obtain $M=1$, because the initial states are already prepared in the eigenstate $X_L = 1$.
  One has to note that this method only works if the qubit states are in $\ket{+}$; otherwise there will be a nonzero chance to obtain the measurement output $M=-1$. The post-measurement state of that output is not the correct entangled state and cannot be used anymore. This is the reason why the ICM representation is incompatible and has to be replaced later on by an inverted ICM representation.

  Thus, one can see that any state which can be prepared with $\ket{+}$ and $\ket{0}$ and CNOTs can also be prepared using split and merge operations. Here, one does not need the logical ancilla qubit that would be required when applying the full CNOT circuit. States that can be prepared using $\ket{0}$ and $\ket{+}$ inputs and CNOT gates are subsets of states known as Calderbank-Shor-Steane stabilized states~\cite{Nielsen2011}. These stabilized states are characterized as having two sets of stabilizers, one consisting purely of $X$ operators and one purely of $Z$.
  As a further remark, this derivation can similarly be performed by using multi-target CNOTs.

\section{\label{sec:ICMRep} ICM Representation}
\begin{figure}
  \includegraphics[width=0.9\columnwidth]{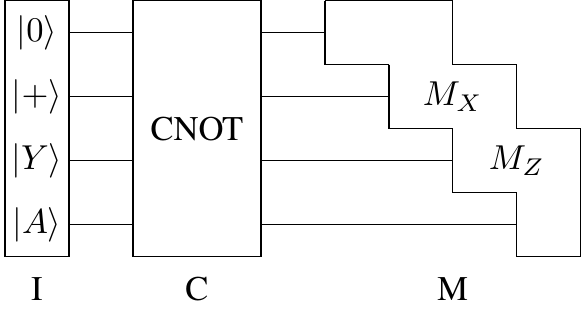}
  \caption{\label{fig:ICMexample} The circuit here is divided into three distinct steps. First in the initialization stage (I), the qubits will be initialized in one of the following states: $\ket{0}$, $\ket{+}$, $\ket{Y}$ or $\ket{A}$.
  In the second step (C), a layer of CNOTs is applied to these states. In the last step, staggered measurements (M) in the $X$- or $Z$-basis are applied, where each basis is chosen depending on the previous measurement results.}
  \end{figure}

  The compilation procedure presented in this paper requires an input circuit that is given by a fault tolerant circuit with a reduced gate set (stage 3 of the proposed design stack in the introduction). To this end we will introduce the ICM model that was devised in reference~\cite{PPND15}. Circuits in the ICM model perform first an initialization of all qubits then an array of CNOTs is applied and measurements are performed in the end. A sample schematic of this is shown in Figure~\ref{fig:ICMexample}. Any higher-level circuit can be rewritten in this form, which includes all necessary ancilliary protocols for fault-tolerant quantum error-correction. A required gate set for the ICM representation is given by the gates: CNOT, $H$, $P$ and $T$.
  The main idea of the ICM model is to use quantum teleportation to implement the operators $H$, $P$ and $T$ at the cost of introducing specially prepared ancilla qubits. This allows an arbitrary circuit to be constructed with a {\em deterministic\/} number of logical qubits and array of CNOT gates. The left parts of Figures~\ref{fig:inversetranslation} and~\ref{fig:inversetranslation2} show two teleportation circuits. The teleportation-based circuits are probabilistic, such that correctional gates might need to be applied after the measurement of the ancilla qubit. This would require a dynamically changing quantum circuit, depending on the measurement result of the ancilla qubits during teleportation. Yet the ICM model shows a way to mitigate this problem by introducing more ancilla qubits and performing selective target teleportation and selective source teleportation circuits~\cite{F12+}, such that the ICM model achieves a deterministic gate array for fault-tolerant, error-corrected computation.

  This form of ICM requires a simulation of CNOTs using splits and merges, which requires further ancillary regions. This is due to the requirement to initialize some logical qubits in the $\ket{A}$ or $\ket{Y}$ state for teleported $T$- and $P$-gates. We want to obtain an algorithmically-specific entangled state with as few ancillas as possible. Thus, we devise a similar representation to ICM, which uses the merge and split operation natively. An example has already been given in section~\ref{chap:merge_split}, but we will now formalize this such that arbitrary circuits can be implemented.

\subsection{Inverse ICM}
\begin{figure}
  \includegraphics[width=0.9\columnwidth]{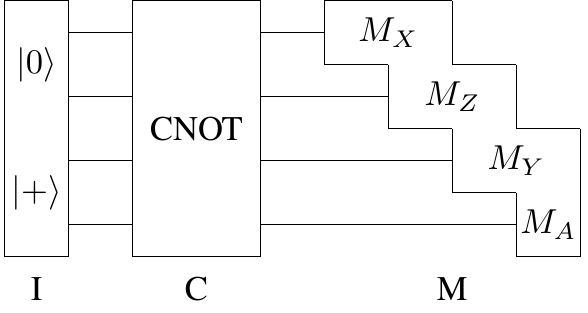}
  \caption{\label{fig:invICMexample} Similarly to the ICM circuit, this inverted ICM circuit has three steps: initialization (I), CNOT (C) and measurement (M). This time, the initialization step only prepares states in $\ket{0}$ and $\ket{+}$. This comes at the expense that measurements have to be performed additionally in the $Y$- and $A$-basis. This form is now compatible with the split and merge operations from lattice surgery.}
  \end{figure}

\begin{figure}
    \subfloat{
    \includegraphics[width=0.4\columnwidth]{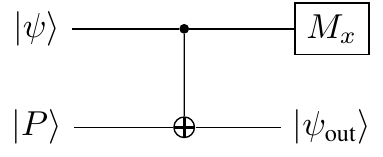}
    }
    \hfill
    \subfloat{
    \includegraphics[width=0.4\columnwidth]{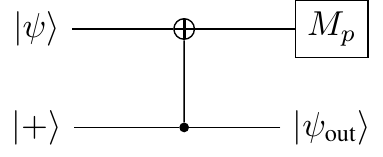}
    }
    \caption{\label{fig:inversetranslation} The circuit on the left is one part of the ICM circuit whereas the circuit on the right constitutes the corresponding inverted ICM model. These two circuits are identical for any state of the form ${\ket{\theta} =\frac{1}{\sqrt{2}}\left( \ket{0} + e^{i\theta} \ket{1}\right)}$ and a general $\ket{\psi}$, which is mapped to $\ket{\psi_{\text{out}}} = R_z(\theta)\ket{\psi}$.}
  \end{figure}

\begin{figure}
    \subfloat{
    \includegraphics[width=0.4\columnwidth]{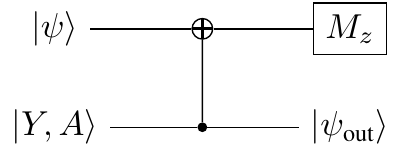}
    }
    \hfill
    \subfloat{
    \includegraphics[width=0.4\columnwidth]{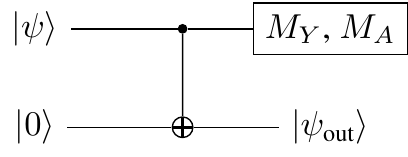}
    }
    \caption{\label{fig:inversetranslation2} The ICM circuit on the left is again reformulated to the inverted circuit on the right. Depending on which state $\ket{Y}$ or $\ket{A}$ was given in the original circuit, a different correction needs to be applied. For the $\ket{Y}$ state a $X$-correction is required, while for the $\ket{A}$ state a $SX$ error occurs, which can be corrected without violating any requirements of the inverted ICM formulation.  The $Y$- and $A$-basis measurements are achieved via a $P$- or $T$-gate, followed by an $X$-basis measurement.}
  \end{figure}

  Using the shorthand implementation of multi-target CNOTs, a merge is only guaranteed to result in the same transformation as a full CNOT gate if the initial qubits are given in the states $\ket{+}$ or $\ket{0}$. Thus, the ICM formulation cannot be used as is and an inverted model is required, where the initialization step only prepares these two states (Figure \ref{fig:invICMexample}). A translation to this form has already been described in~\cite{RotatedICM} and will be outlined here. For this discussion we will use a phase state ${\ket{\theta} =\frac{1}{\sqrt{2}}\left( \ket{0} + e^{i\theta} \ket{1}\right)}$ and an arbitrary state $\ket{\psi} = \alpha \ket{0} + \beta \ket{1}$ to show equivalences, which then apply for $\theta = \{\pi/4,\pi/2\}$, for the $\ket{A}$ and $\ket{Y}$ states needed in a fault-tolerant implementation. In order to prove the equivalence of ICM and inverted ICM, only two cases need to be considered.
  These cases are given in Figure~\ref{fig:inversetranslation} and Figure~\ref{fig:inversetranslation2}.
  An inverted ICM circuit can be constructed by combining these sub-circuits, such that initialization is only in the $\ket{0}$ or $\ket{+}$ state and $P$- and $T$-gates are realized through rotated basis measurements.

  The first equivalence is shown in Figure~\ref{fig:inversetranslation}. Depending on the measurement outcome, the output state of the second qubit is either given by ${\ket{\psi_\text{out}}=\left(\alpha + e^{i\theta}\beta \right)\ket{0} +\left(e^{i\theta}\alpha + \beta\right)\ket{1}}$
  or ${\ket{\psi_\text{out}}=\left(\alpha - e^{i\theta} \beta \right)\ket{0}+ \left(e^{i\theta} \alpha - \beta\right)\ket{1}}$. This holds true for both circuits shown.

  The equivalence of Figure~\ref{fig:inversetranslation2} is not as straightforward as for the one before because the output state needs to be corrected. For the following argument we calculate the effect of the circuits only for the case of an $\ket{A}$ state. The $\ket{Y}$ state can be calculated analogously.
  The output states for the ICM circuit (left) are either $\ket{\psi_\text{out}}= \alpha\ket{0} + e^{\frac{\pi i}{4}} \beta \ket{1}$, for a $\ket{0}$ measurement, or $\ket{\psi_\text{out}}= \beta \ket{0} + e^{\frac{\pi i}{4}} \alpha \ket{1}$, when the measurement is $\ket{1}$.
  That is, if we measure $\ket{1}$, a $T^{\dagger}$-gate is applied instead of a $T$-gate.
  Thus the $P$-gate needs to be applied in order to correct for this error, as $PT^{\dagger} = T$. This error cannot be tracked and needs to be applied using the previously mentioned selective destination teleportation algorithms~\cite{F12+}. After that, the qubit is in a state of $X Z \ket{\psi}$, where the Pauli operators can be tracked classically. The $P$-gate has a similar correction, but as this is a simple $Z$-gate, $ZP^{\dagger} = P$, this can be classically tracked without any further correction.

  The inverted ICM measurement has the advantage that this correction is not needed. For a $\ket{0}$ measurement the state is the same as for the ICM circuit. However, if the inverted ICM measurement returns a $\ket{1}$, the output state will be ${\ket{\psi_\text{out}} = \alpha \ket{0} - e^{\frac{i\pi}{4}} \beta \ket{1}}$ for the $T$-gate and ${\ket{\psi_\text{out}} = \alpha \ket{0} - i \beta \ket{1}}$ for the $P$-gate, which only requires a Pauli $Z$-correction.


  These circuits can now be stacked together such that any circuit given in ICM format can be translated by removing all the rotated state initializations and replacing these by rotated measurements.
  However, the implementation of such measurements are not fault-tolerantly protected in the surface code and basis transformations given by $T$- and $P$-gates need to be performed. For these gates, state injection is needed. Furthermore, a $T$-basis transformation is probabilistic and requires further correctional $\ket{Y}$ states. 

\section{\label{sec:IC} Implementation of I and C}
  Using the inverted ICM model, we can now derive an algorithm that prepares the required {\em logically\/} entangled states using purely the native split/merge operations in the lattice surgery model. The (I) and (C) parts of the circuit create a specific Calderbank-Shor-Steane stabilized state that reflects the overlying algorithm, after which rotated logical measurements are performed (in an identical way to the original description of cluster state quantum computation~\cite{MBQC_oneway}, were all Pauli measurements are made on a 2D cluster state resource, defining an algorithmically-specific graph state consisting of rotated basis measurements and feedforward).
  Since the inverse ICM model is a universal model for fault-tolerant, error-corrected quantum computation, so is the presented model, with the proposed translation. The translated circuit relies on smooth split operations followed by rough merge operations. After this, the qubits are in an algorithmically-specific encoded state, ready for measurements and feedforward to be performed.

\subsection{Classical Algorithm}
\begin{figure}
    \subfloat{
    \includegraphics[width=0.45\columnwidth]{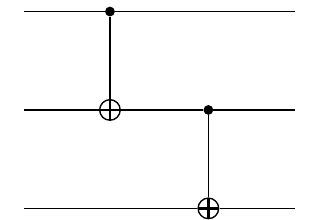}
    }
    \subfloat{
    \includegraphics[width=0.45\columnwidth]{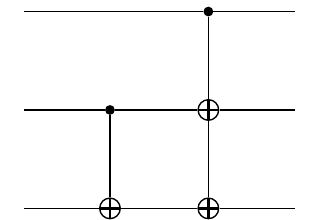}
    }
    \caption{\label{fig:swapCNOT1} CNOT commutation relations. The circuit on the left has the same effect as the circuit on the right, such that this identity can be used to reformulate circuits.}
  \end{figure}

\begin{figure}
    \subfloat{
    \includegraphics[width=0.45\columnwidth]{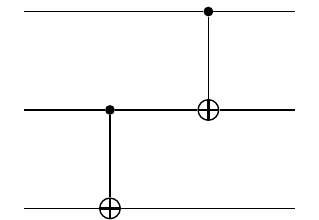}
    }
    \subfloat{
    \includegraphics[width=0.45\columnwidth]{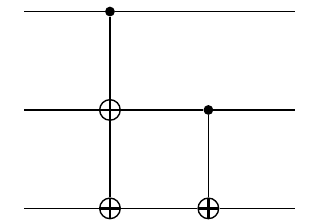}
    }
  \caption{\label{fig:swapCNOT2} CNOT commutation relations, which can be used to reformulate circuits and achieve the form required by lattice surgery.}
  \end{figure}

  The classical algorithm given as additional material can be used to translate the circuit to a representation that can be implemented by the lattice surgery model. It reduces the number of (multi-target) CNOTs to a minimum.
  The algorithm relies on three simple circuit-modification rules:
  \begin{enumerate}
    \item Any CNOT that targets an unentangled $\ket{+}$ has no effect, as this is the eigenstate with eigenvalue $1$.
    \item A CNOT whose control qubit is in $\ket{0}$ does not have any effect.
    \item A CNOT with the same target and control qubits as its neighbor acts like the identity.
  \end{enumerate}
  Furthermore, we use the commutation relations between different non-commuting CNOTs shown in Figure~\ref{fig:swapCNOT1} and~\ref{fig:swapCNOT2}.

  Using the rules above, the algorithm now proceeds as follows: In a first step, all CNOTs that target a qubit initialized to $\ket{+}$ are permuted to the beginning and using rule 1 the CNOTs are deleted. Then all CNOTs with a target qubit that initialized to the $\ket{0}$ state are also moved to the front individually and are deleted using rule 2. Due to the commutation relations given in Figure~\ref{fig:swapCNOT1} and Figure~\ref{fig:swapCNOT2} new CNOTs are created, but their control qubit is not located on a $\ket{0}$ state. This reduces the circuit to one where each CNOT has a control qubit initialized to the $\ket{+}$ state at the beginning. Many redundant CNOTs were introduced during the permutation actions, which need to be cleaned up. Thus in a final step all CNOTs operating on the same pair of qubits are moved together and are annihilated using rule 3. Because of rule 1 the control qubit of any CNOT is not targeted by any other CNOT such that each CNOT commutes with the others. Thus the cleanup can be performed easily.

\subsection{Lattice surgery translation}
  Now any circuit can be translated to a circuit that has some number of multi-target CNOTs, whose control qubits are in the $\ket{+}$ state. A circuit of this form is easy to translate to the lattice surgery model. Its procedure requires four distinct steps. First, all the initial $\ket{+}$ states are prepared. Using smooth splits these qubits will, in a second step, be split into independent groups of entangled {\em logical\/} qubits, depending on the number of target qubits each CNOT has. In order to connect these independent blocks one will (in the fourth step) merge corresponding qubits from each independent block. This merging operation has to occur on neighboring blocks in the lattice surgery model, such that in an intermediate third step one needs to shrink or grow the size of the surface code patches or move them to another location.

\begin{figure}
  \includegraphics[width=\columnwidth]{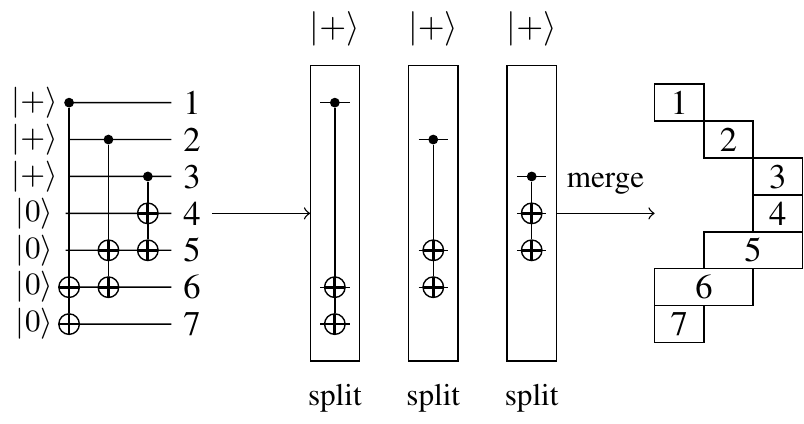}
  \caption{\label{fig:LStranslation} This illustrates what operations have to be implemented to perform the I and C parts of any inverted ICM circuit. First, each CNOT gets its own vertical patch of surface code, initialized to $\ket{+}$. Using smooth splits these CNOTs are split into their individual qubits. Common qubits among the different CNOTs are merged, such that the output of the circuit looks akin to the right configuration. Each number here indicates which qubit from the original circuit is encoded in which patch of surface code. In this case no effort was spent on optimization which reduces the number of steps to three: preparation, smooth splits and rough merges.}
\end{figure}

  An example for this translation is shown in Figure~\ref{fig:LStranslation}. Here a circuit with three multi-target CNOTs is translated to the lattice surgery model with no optimization. First, a vertical patch of the surface code is initialized to $\ket{+}$ for each multi-target CNOT. The first multi-target CNOT consists of qubits 1, 6 and 7. Thus, the first patch is split into 3 patches, which can be labeled 1, 6 and 7 indicating which patch corresponds to which qubit in the circuit. It should be noted, that the ordering of these labels is arbitrary, because, so far, each patch encodes the same state. However, multiple patches corresponding to the same qubit exist (labeled by the same number), such that merges are necessary between them. This unoptimized illustration assigns each qubit an individual row, such that merges are always possible. This makes it clear that the space-cost in this representation is upper-bounded by the number of logical qubits times the number of multi-target CNOT patches. For the circuits, later on we will optimize the placement of these patches manually.

  For a large quantum circuit, the placement of these logical regions within the overall computer and how they are split/merged and shifted around will dictate overall resource costs in terms of physical qubits and computational time. We illustrate three explicit examples for state distillation circuits that are extensively used in surface code computation, but the more general problem of scheduling and resource optimization using this technique is relegated to future work. A more detailed analysis, combined with recent results reducing physical resources in the lattice surgery model~\cite{Campbelldistill,DIP16,NSM16} is anticipated to result in better performance than topological braiding (Note: the optimization problem has not yet been solved for braiding-based logic, so an ultimate comparison will only be fair when both problems are finally solved).

\section{\label{sec:stabilmatrix} Stabilizer matrix}
  The inverted ICM model is designed to create an algorithmically-specific Calderbank-Shor-Steane-stabilized state~\cite{CSS,Nielsen2011} that is compatible with fault-tolerant and error-correction protocols. The split and merge operations in lattice surgery can also be re-written in terms of stabilizer transformations, which allows us to link the two in a straightforward manner.
  \subsection{Merge Operation}
  During a merge between two encoded qubits, the number of encoded degrees of freedom decreases by one. For a rough merge the rules are given as follows
  \begin{align*}
    \begin{bmatrix}X&I\end{bmatrix} &\rightarrow \begin{bmatrix}X\end{bmatrix} \\
    \begin{bmatrix}I&X\end{bmatrix} &\rightarrow \begin{bmatrix}X\end{bmatrix} \\
    \begin{bmatrix}I&I\end{bmatrix} &\rightarrow \begin{bmatrix}I\end{bmatrix}.
  \end{align*}
  For the logical $Z$-operators, this action is not as easy, since it might be that stabilizers need to be merged. If there are no $Z$ stabilizers acting on the two qubits that partake in the merge operation, nothing has to be done. If only one of the qubits is affected by a $Z$ stabilizer, the post-merge rule for the merged qubit is given by
  \begin{align*}
  \begin{bmatrix}Z&I\end{bmatrix}\rightarrow \begin{bmatrix}Z\end{bmatrix}   \qquad \qquad
  \begin{bmatrix}I&Z\end{bmatrix}\rightarrow \begin{bmatrix}Z\end{bmatrix}.
  \end{align*}
  For a general stabilizer matrix, many $Z$ stabilizers might act on the same qubit. Since any linear combination of these stabilizers gives an equivalent stabilizer description, one can add and subtract stabilizer rows without changing the behavior of the stabilizer. If the two qubits are acted upon using more than two $Z$ stabilizers, one can always find a representation in which only one $Z$ operator exists in each column of the merging qubits.
  The two stabilizer rows are merged using the following rules:
  \begin{align*}
    &\begin{bmatrix}Z\\I\end{bmatrix} \rightarrow \begin{bmatrix}Z \end{bmatrix}
      \qquad \qquad
    \begin{bmatrix}I\\Z\end{bmatrix} \rightarrow \begin{bmatrix}Z \end{bmatrix} \\
    &\begin{bmatrix}I\\I\end{bmatrix} \rightarrow \begin{bmatrix}I \end{bmatrix}
      \qquad \qquad \; \,
    \begin{bmatrix}Z\\Z\end{bmatrix} \rightarrow \begin{bmatrix}I \end{bmatrix}
  \end{align*}

  \subsection{Split Operation}
  The physical result of a general split operation is given in Equation~(\ref{eq:roughsplit}) for a rough split and Equation~(\ref{eq:smoothsplit}) for a smooth split. Due to the creation of a new encoded qubit during the process of the split operation, a new column has to be created in the stabilizer matrix. Since the created qubit is linked to the pre-split state, a new stabilizer row has to be created as well.
  This new stabilizer is given by $ZZ$, for a smooth split at the position of the two affected qubit positions. Furthermore, any pre-split stabilizer will transform using the following rules
  \begin{equation*}
    \begin{bmatrix}Z\end{bmatrix} \rightarrow \begin{bmatrix}Z&I \end{bmatrix}
      \qquad
    \begin{bmatrix}X\end{bmatrix} \rightarrow \begin{bmatrix}X&X\end{bmatrix}
      \qquad
    \begin{bmatrix}I\end{bmatrix} \rightarrow \begin{bmatrix}I&I \end{bmatrix}
  \end{equation*}
  for a smooth split. Exchanging $X$ with $Z$, one will get the relations for a rough merge.

  \subsection{Example algorithm}
  The resulting state of the example circuit in Figure~\ref{fig:CNOTex} will give the stabilizer matrix:
  \begin{equation*}
    \begin{bmatrix}
      Z&Z&Z\\
      X&X&I\\
      I&X&X
    \end{bmatrix}
  \end{equation*}
  Using the rules described above we will derive the stabilizer matrix using lattice surgery moves. First, we start in a state with two qubits each initialized individually into $X$-eigenstates. Then both an additional column for the additional qubit and an additional line for the stabilizer are created.
  \begin{equation*}
    \begin{bmatrix}
      X&I\\
      I&X
    \end{bmatrix}
    \rightarrow
    \begin{bmatrix}
      X&&I\\
      & & \\
      I& &X
    \end{bmatrix}
    \rightarrow
    \begin{bmatrix}
      X&X&I\\
      Z&Z&I\\
      I&I&X
    \end{bmatrix}
  \end{equation*}
  After the same steps are performed on the other qubit, the middle qubits are merged.
  \begin{equation*}
    \begin{bmatrix}
      X&X&I&I\\
      Z&Z&I&I\\
      I&I&X&X\\
      I&I&Z&Z
    \end{bmatrix}
   \rightarrow
   \begin{bmatrix}
     Z&Z&Z\\
     X&X&I\\
     I&X&X
   \end{bmatrix}
  \end{equation*}
  The previously mentioned procedures can now perform the (I) and (C) steps of the inverted ICM model using the inherent merge and split operations of lattice surgery. The remaining part is to illustrate how the rotated basis measurements of the inverted ICM model can be performed in lattice surgery.

\section{\label{sec:measurement} Measurement step}
  Having prepared an entangled state using I and C of the inverted ICM representation, measurements need to be performed. But for the 2D surface-code patches used in this paper, fault-tolerant measurements in arbitrary bases are not possible. This requires the application of basis transformations which rely on the injection of magic states. The main work during the measurement step consists of the preparation and application of these basis transformations, which we direct our focus to in the following.

  The method described in the following can be applied to any phase gate, but in this paper we will focus only on the $P$- and the non-Clifford $T$-gate.
  In surface code implementations these gates are hard to implement, as QEC does not support their immediate application. Thus the common technique used so far is state-injection, where magic states are prepared and used to perform such gates. These magic states are given by
  \begin{align}
    \ket{Y} &= \frac{1}{\sqrt{2}} \left(\ket{0} + i \ket{1}\right) \\
    \ket{A} &=  \frac{1}{\sqrt{2}} \left(\ket{0} + e^{i\frac{\pi}{4}} \ket{1} \right).
  \end{align}
  A magic state is obtained by manipulating a single physical qubit and then encoding it in the error-correcting framework.
  Since this process is not fault-tolerant, the error on the resultant logical state is dominated by the physical preparation of the ancilla and its encoding.
  To purify the state, a process called state distillation is required~\cite{BK05+}. The main effort in the application of $P$- and $T$-gates is spent distilling magic states, such that an efficient way to perform the distillation procedure will give an efficient $P$- or $T$-gate. At the end of this paper the performance of these distillation algorithms will be compared to braiding using the translation proposed in this paper.

  If a suitable (clean) ancilla qubit is available, the $P$- or $T$-gates can be applied by the circuits given in~\ref{fig:inversetranslation} and~\ref{fig:inversetranslation2}. In fact, any quantum computer that aims at outperforming classical hardware has to rely on non-Clifford gates, since any circuit without these can be simulated efficiently using classical hardware~\cite{efficientsim}. This makes a resource-friendly application of the $T$-gate crucial.

  One should note that the implementation of a $P$-gate requires one ancilla state set to $\ket{Y}$. Since these are teleportation-based protocols, a measurement has to be performed to apply the operator. With this measurement, the magic state resource gets destroyed for both the $P$- and the $T$-gate. The destruction of $\ket{Y}$-state ancillas can be avoided using a technique~\cite{A07,J13+} which requires direct Hadamard operations on a planar code, which is possible using code-deformed techniques~\cite{F12}. However, we do not explicitly consider this here and instead look at distillation circuit constructions for the $\ket{Y}$ state.

\subsection{Special Cases Merge}
  Here we will explicitly calculate the post-merge state of a smooth merge between an arbitrary phase qubit ${\ket{P} = \ket{0}+ p \cdot \ket{1}}$ and a general qubit ${\ket{\phi}=\alpha \ket{0} + \beta\ket{1}}$. With this we will show that both the $P$- and $T$-gates can be efficiently implemented using a single smooth merge operation and only require as a prerequisite the magic state $\ket{Y}$ or $\ket{A}$.
  In the conjugate basis, the states are given by
  \begin{align*}
  \ket{P} &= \left( \frac{1+p}{2} \right) \ket{+} + \left( \frac{1-p}{2} \right) \ket{-} \\
  \ket{\phi} &=\left( \frac{\alpha + \beta}{2} \right) \ket{+} + \left( \frac{\alpha - \beta}{2} \right) \ket{-},
  \end{align*}
  such that we can insert this into the post-merge state:
  \begin{equation*}
    \ket{A} \encircle{M}_s \ket{\phi} = \left( \frac{\alpha+\beta}{2} \right) \ket{P} + {(-1)}^M \left( \frac{\alpha - \beta}{2} \right) \ket{\overline{P}}
  \end{equation*}
  If the measurement result was $M=0$, then this evaluates to:
  \begin{align*}
    \ket{A} \encircle{M}_s \ket{\phi} &=\left( \frac{\alpha}{2} + \frac{\beta p}{2} \right) \ket{+} +\left( \frac{\alpha}{2} - \frac{\beta p}{2}\right) \ket{-} \\
    &=\alpha \ket{0} + \beta \, p \ket{1}
  \end{align*}
  If the measurement was $M=1$, then the state will be given by:
  \begin{align*}
    \ket{A} \encircle{M}_s \ket{\phi} &=\left( \frac{\alpha p}{2} + \frac{\beta}{2} \right) \ket{+} +\left( -\frac{\alpha p}{2} + \frac{\beta}{2}\right) \ket{-} \\
    &=\beta \ket{0} + \alpha \, p \ket{1}
  \end{align*}
  Setting $p=i$, we obtain an implementation for the $P$-gate. If $M=0$, the effect of this merge operation already coincides with $P\ket{\psi}$. But in this case of $M=1$, a Pauli-$X$ and Pauli-$Z$ correction have to be applied.

  Moreover, for an implementation of the $T$-gate we require $p=e^{i \frac{\pi}{4}}$. Again, if $M=0$ no corrections are needed. Yet for $M=1$, instead of the $T$-gate the operator $X T^\dagger$ was applied. This can be corrected by first performing a Pauli-$X$ operator and then a $P$-gate. The $P$-gate correction needs to be physically applied and cannot be tracked, and a selective source/destination teleportation algorithm has to be employed. Thus this lattice surgery implementation has the same drawbacks as the teleportation circuits when we perform the rotated-basis measurements.

  One should note that the application of this merge can be thought of a basis transform which is needed in the measurement step of the inverted ICM model, where measurements in the $A$- and $Y$-basis are needed.

  \subsection{Encoding}
  The last part needed for universal computation is how to encode one of the magic states in a lattice surgery patch. For completeness, a summary of the required steps from the original papers~\cite{Horsman2012,encoding_magic,encoding_magic_2} is given in the following.

  First a physical qubit is prepared in the desired state. All the other qubits of the error-correcting surface code need to be initialized to the state $\ket{0}$.

  Using the CNOTs that are required for the measurement of the planar code stabilizers, this state can be transformed to a superposition state involving the original data qubits and the syndrome qubits immediately above and below it. These syndrome qubits can now be swapped with the data qubits on the opposite side of the original qubit. This results in a state where a vertical line of three data qubits in the error-correcting surface code are initialized to the state
  \begin{equation}
    \ket{\psi} = \alpha \ket{000} + \beta \ket{111}.
  \end{equation}
  Now the stabilizers for a distance-three error-correcting surface patch around these specially prepared qubits can be turned on, and thus an error-corrected state is obtained. The distance of these states can be increased by performing merge operations with neighboring regions. Here, the neighboring qubits need to be initialized to $\ket{0}$ again.

\section{\label{sec:example} Examples}
  This concludes the description of the compilation procedure. Any circuit that has been translated to the inverse ICM representation can now be implemented in the planar code using efficient operations that the 2D lattice of surface code patches enables. After initialization, the CNOT step is entirely implemented using merge and split operations; and the last step of measurements uses state distillation and injection as described in the section~\ref{sec:measurement}. In the following, we provide a translation to lattice surgery for three common distillation protocols: the 7-qubit Steane code, the 15-qubit Reed-Muller code, and finally a Bravyi-Haah distillation code.
  For all of these codes we provide a hand-optimized placement of the surface code patches and estimate their resource requirements.

\subsection{Steane code for $\ket{Y}$-state-distillation}
  \begin{figure}
    \centering
    \includegraphics[width=\columnwidth]{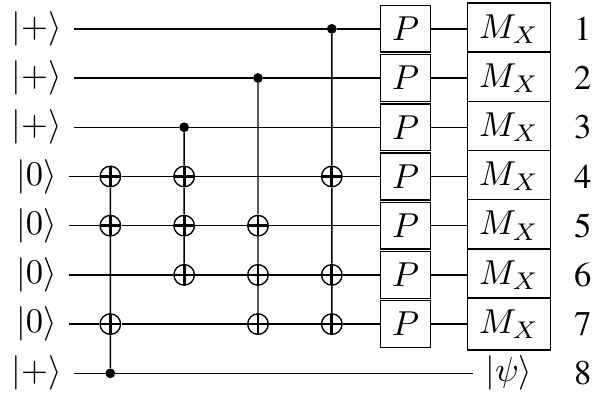}
    \caption{\label{fig:Steane} This circuit is the Steane code, to be used for the distillation of $\ket{Y}$ states. This is an iterative procedure where the error-prone $\ket{Y}$ are used during the application of the $P$-gates. The numbering used in this circuit coincides with the numbering of the algorithm given in Figure~\ref{fig:algorithm}.}
  \end{figure}

  \begin{figure}
    \subfloat[\label{fig:S_init}For each multi target CNOT a vertical block of surface code is initialized to $\ket{+}$. The height of these blocks is determined by the number of target qubits and how the circuit is optimized.]{
    \includegraphics[width=0.4\columnwidth]{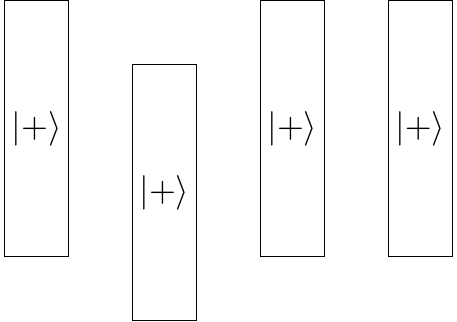}
  }
  \hfill
  \subfloat[\label{fig:S_split}Each block is now separated into its qubits using smooth splits.]{
  \includegraphics[width=0.4\columnwidth]{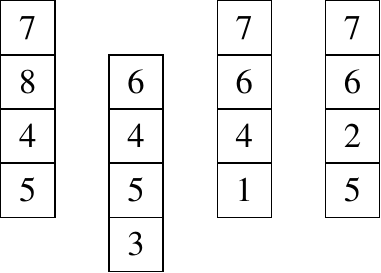}
  }
  \vspace{0.4cm}
  \subfloat[\label{fig:S_merge}A rough merge connects the same target qubits of different CNOTs.]{
  \includegraphics[width=0.4\columnwidth]{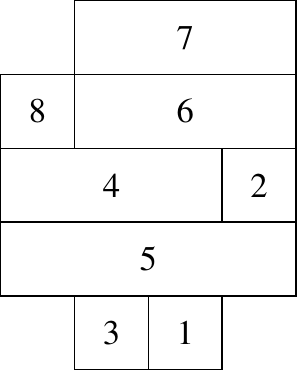}
  }
  \hfill
  \subfloat[\label{fig:S_shrink}A shrink operation reduces the size of each patch to unitary in order to make space for the $\ket{Y}$ states.]{
  \includegraphics[width=0.4\columnwidth]{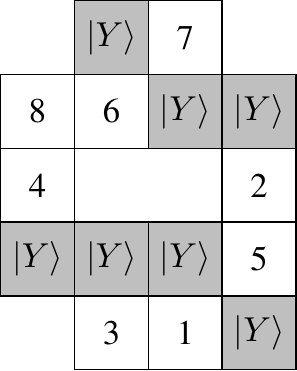}
  }
  \vspace{0.4cm}
  \subfloat[\label{fig:S_final}A smooth merge between a $\ket{Y}$-state and another qubit performs injection, such that the $Z$-gate is applied.]{
  \includegraphics[width=0.4\columnwidth]{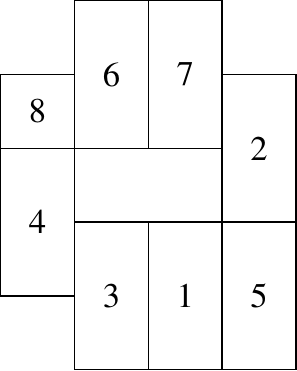}
  }
  \vspace{0.2cm}
  \caption{\label{fig:algorithm} This figure shows the steps needed to perform lattice surgery. The numbers indicate which patch contributes to which qubit of the circuit in Fig.~\ref{fig:Steane}. Since merges and splits do not conserve the overall number of qubits, many patches might contribute to the same qubit at intermediate steps.}
  \end{figure}

  As a first example, we show how the state distillation algorithm for the $P$-gate can be translated to the lattice surgery model. The underlying error-correcting code for the distillation is a 7-qubit Steane code given in Figure~\ref{fig:Steane}. One can see that this is already given in the inverse ICM format, if one replaces the $P$-gates and $X$-basis measurements by a rotated measurement.

  For this algorithm, eight encoded qubit regions are needed, which are in the beginning prepared in an entagled state using the multi-target CNOT operations. This distillation circuit now applies to seven encoded qubits an error-prone $P$-gate. After that, the ancilla qubits are measured and depending on the outcome this circuit will have a distilled version as output $\ket{\psi} = \ket{Y_{k+1}}$, where $(k+1)$ denotes the output as a distilled $\ket{Y}$ state at $(k+1)$ levels of concatenation. This algorithm can be used recursively until the desired precision is reached.

 At the measurement stage, we are checking the eigenvalues of the three stabilizers,
  \begin{align*}
    S_1 &= M_X^1 M_X^4 M_X^6 M_X^7\\
    S_2 &= M_X^2 M_X^5 M_X^6 M_X^7\\
    S_3 &= M_X^3 M_X^4 M_X^5 M_X^6.
  \end{align*}
  Only if all of these stabilizers return the trivial syndrome, then the distillation procedure works. Otherwise, the state is discarded and another distillation run has to be performed. Furthermore, if the product of all measurements is ${M_X^1 \cdots M_X^7=1}$, then the output state is given by $\ket{\psi} = \sigma_z \ket{Y}$, whose $Z$-error needs to be tracked.

  Since the multi-qubit CNOTs of the Steane code prepare the system in a Calderbank-Shor-Steane-stabilized state, one can easily translate this to split operations in the lattice surgery model. No further classical translation is required. Using the method described before, each CNOT corresponds to an instance of smooth splits, which will then be connected by rough merges. The placement of the individual qubits can be treated as an optimization problem in order to place corresponding qubits close to each other.

  A method to proceed is given in Figure~\ref{fig:algorithm}. The procedure starts with the initialization of four encoded regions to the state $\ket{+}^{\otimes 4}$. Because smooth splits will be performed, these four regions have sufficient size to accommodate four encoded regions each. Afterwards, in Figure~\ref{fig:S_split}, the smooth splits are performed creating all qubits that are needed in the original circuit given in Figure~\ref{fig:Steane}.
  In the next step, qubit one will be moved one space down and the leftmost qubit 7 will be moved to the right. After that, all patches contributing to the same qubit are merged using a rough merge. The result is visualized in Figure~\ref{fig:S_merge}.

  At this point, we have prepared the entangled state between eight logically-encoded regions that reflect both the initialization and CNOT parts of the distillation circuit.  The remaining operation is to perform each individual logical $P$-gate on qubit regions one to seven and measure them out in the $X$-basis. For each of these qubits we need to introduce ancillary $\ket{Y}$ states.  Without loss of generality, we assume that this is a level-one concatenated circuit, end hence we need to state-inject seven physical $\ket{Y}$ states and encode them into encoded regions. To free up lattice space, we first shrink qubit regions seven, six, four, and five and use the resulting lattice space to inject and encode $\ket{Y}$ states which are adjacent to the encoded regions that are needed in Figure~\ref{fig:Steane}.
  Finally, smooth merges are performed and the resulting qubits will be measured such that only logical qubit 8 remains, which is our output, with the rest of the code space now free to be used in subsequent circuits.

  Using this procedure recursively will exponentially decrease the errors associated with the magic state. Therefore, one can obtain an arbitrary precise $Y$-state and thus the application of a arbitrarily precise $P$-gate is possible. However, higher levels of such concatenations need $\ket{Y}$-states calculated before. The transportation of these states complicates the geometry and further research is required.

\subsection{Efficiency of the distillation}
  The total spatial requirements for one distillation run using the algorithm described above are given by $5\cdot4$ patches that encode a logical qubit each. If a rotated lattice is used, one will need $d^2$ data qubits, and $(d^2-1)$ syndrome qubits~\cite{Horsman2012} for a distance $d$ surface code. This results in $2d^2$ physical qubits to leading order. The time requirements are given by $d$-cycles for the initialization of the states $\ket{+}$. Performing all the smooth splits in Figure~\ref{fig:S_split} will need $d$-cycles. The movement and merge operations need $d$-cycles each. Shrinking the qubits also needs $d$-cycles, while creating the injected $\ket{Y}$ states needs at least $d$-cycles. The final smooth merges take again $d$-cycles, such that the total time requirements sum to
  \begin{equation*}
    t = 7d \qquad \text{cycles}.
  \end{equation*}
  This will give a total requirement of ${2\cdot20\cdot7d^3 = 280 d^3}$ space-time volumes, if a rotated lattice is assumed (it should be noted that one step in time consists of two stabilizer rounds: $Z$-stabilizer and $X$-stabilizer). This performs worse than using the braiding description, which needs a space/time volume of $140d^3$~\cite{FD12}.

\subsection{Stabilizer Matrix Calculation}
  The previously outlined algorithm can be calculated alternatively using the stabilizer matrix formulation with the rules presented before. In the beginning only four encoded qubits, in the state $\ket{+}$, exits, which can be represented as:
  \begin{equation*}
    \begin{bmatrix}
      X &  &  & \\
       & X &  & \\
       &  & X & \\
       &  &  & X\\
    \end{bmatrix}
  \end{equation*}
  Using four smooth splits on each of these qubits will result in a four qubit GHZ-state for each.
  \begin{equation*}
    \left[
    \begin{smallmatrix}
      7&8&4&5& 6&4&5&3& 7&6&4&1& 7&6&2&5 \\
      X&X&X&X& &&&& &&&& &&& \\
      Z&Z&&& &&&& &&&& &&& \\
      &Z&Z&& &&&& &&&& &&& \\
      &&Z&Z& &&&& &&&& &&& \\
      &&&& X&X&X&X& &&&& &&& \\
      &&&& Z&Z&&& &&&& &&& \\
      &&&& &Z&Z&& &&&& &&& \\
      &&&& &&Z&Z& &&&& &&& \\
      &&&& &&&& X&X&X&X& &&& \\
      &&&& &&&& Z&Z&&& &&& \\
      &&&& &&&& &Z&Z&& &&& \\
      &&&& &&&& &&Z&Z& &&& \\
      &&&& &&&& &&&& X&X&X&X \\
      &&&& &&&& &&&& Z&Z&& \\
      &&&& &&&& &&&& &Z&Z& \\
      &&&& &&&& &&&& &&Z&Z \\
      &&&& &&&& &&&& &&&&
    \end{smallmatrix}
    \right]
  \end{equation*}
  In this stabilizer matrix, the numbers correspond to the labeling that was used in Figure~\ref{fig:S_split}. Now one has to proceed using the merge operations. The first merge is performed on the qubit labeled 7:
  \begin{equation*}
  \left[
  \begin{smallmatrix}
    7&8&4&5& 6&4&5&3& 6&4&1& 6&2&5 \\
    X&X&X&X& &&&& &&& && \\
    Z&Z&&& &&&& &&Z& &Z& \\
    &Z&Z&& &&&& &&& && \\
    &&Z&Z& &&&& &&& && \\
    &&&& X&X&X&X& &&& && \\
    &&&& Z&Z&&& &&& && \\
    &&&& &Z&Z&& &&& && \\
    &&&& &&Z&Z& &&& && \\
    X&&&& &&&& X&X&X& && \\
    &&&& &&&& Z&Z&& && \\
    &&&& &&&& &Z&Z& && \\
    X&&&& &&&& &&& X&X&X \\
    &&&& &&&& &&& Z&Z& \\
    &&&& &&&& &&& &Z&Z \\
    &&&& &&&& &&& &&&
  \end{smallmatrix}
  \right]
  \end{equation*}
  Using similar transformations, sorting the numbers and swapping the stabilizer rows, one obtains:
  \begin{equation*}
    \left[
    \begin{smallmatrix}
      1&2&3&4& 5&6&7&8 \\
      &&X&X&X &X&& \\
      &X&&&X &&X&X  \\
      &X&&& X&X&X& \\
      &&&X&X &&X&X \\
      Z&Z&Z&& &Z&& \\
      Z&Z&&& &&Z&Z \\
      Z&&Z&Z& &&&Z \\
      &Z&Z&&Z &&&Z \\
      &&&& &&&
      \end{smallmatrix}
    \right]
  \end{equation*}
  This stabilizer matrix describes the same state as the one given in~\cite{FD12}, which was calculated using the circuit of Figure~\ref{fig:Steane}.
\subsection{Reed-Muller code for $\ket{A}$-state-distillation}
\begin{figure}
  \centering
  \includegraphics[width=0.7\columnwidth]{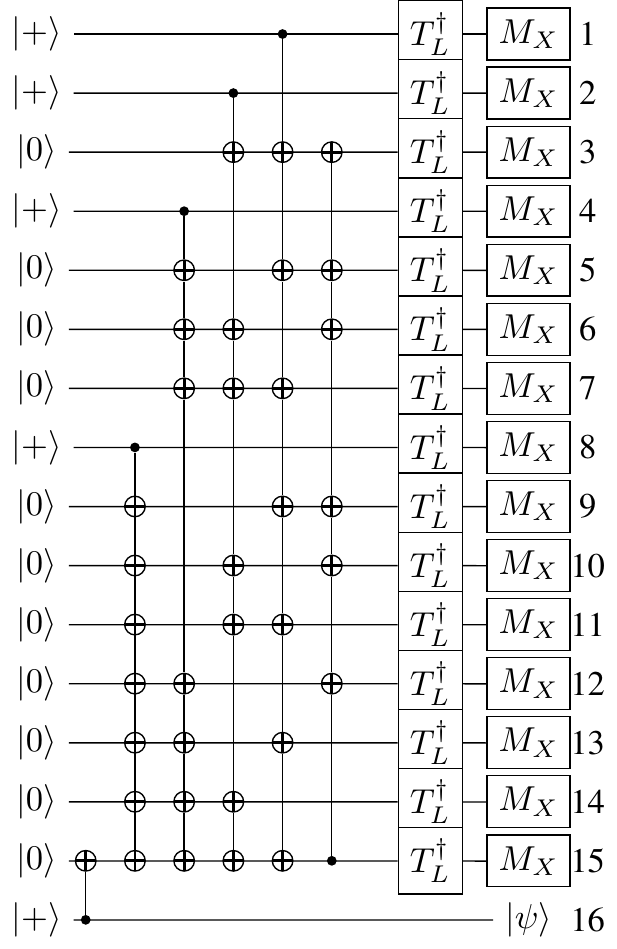}
  \caption{\label{fig:RMuller} This circuit implements a Reed-Muller code which is used to distill $\ket{A}$. However, it is not yet possible to implement this circuit in lattice surgery since the last CNOT has a control qubit on a state that is targeted by other CNOTs. Using the classical algorithm provided in the supplementary, this circuit can be translated to a form that is implementable by lattice surgery.}
\end{figure}

\begin{figure}
  \centering
  \includegraphics[width=0.6\columnwidth]{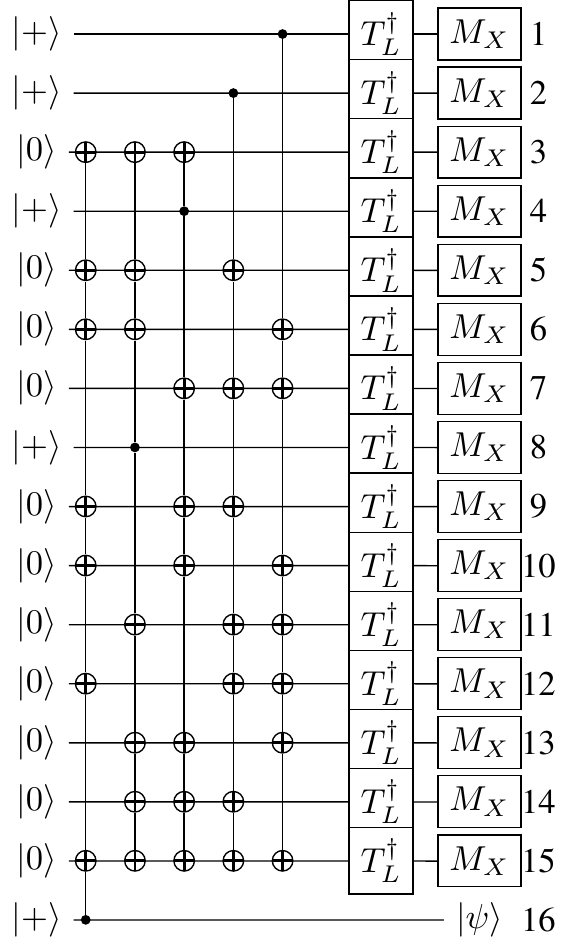}
  \caption{\label{fig:RMuller_opt} The last CNOT of the circuit in Fig.~\ref{fig:RMuller} has been eliminated resulting in an equivalent circuit that can be used with lattice surgery.}
\end{figure}

\begin{figure}
  \includegraphics[width=0.7\columnwidth]{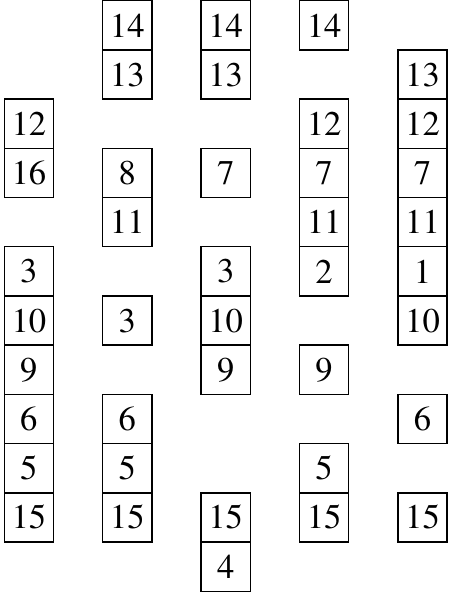}
\caption{\label{fig:RM_layout} This shows the layout after the smooth splits and after their movement to their designated vertical positions. Each column represents one multi target CNOT with the qubits arranged in such a way that in the next step the merge operation can be performed. The empty spaces are initialized to 0, such that all merge operations can be applied in $d$ error-correcting cycles. As a side note, the difference of a $T^\dagger$-gate to a $T$-gate is only a change in corrective gates.}
\end{figure}

\begin{figure}
  \includegraphics[width=0.4\columnwidth]{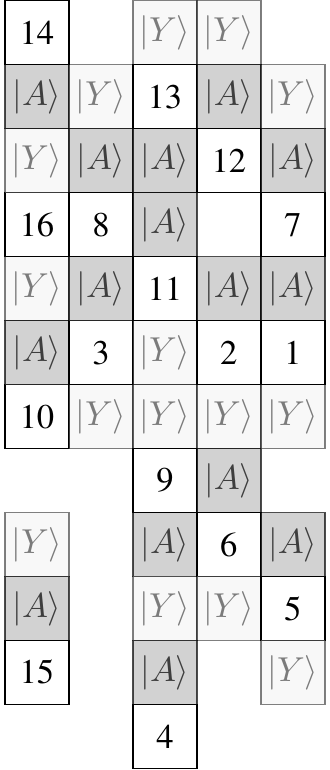}
\caption{\label{fig:RM_injection} After the patches corresponding to the same qubits are merged in Figure~\ref{fig:RM_layout}, excessive patches were shrunk, allowing space for state injection. Here, a possible layout with the correct positioning of $\ket{A}$-states and correctional $\ket{Y}$-states is given. In the next step a smooth merge is performed on every patch except the one labeled 16, which will store the output information.}
\end{figure}

The distillation circuit for $\ket{A}$ is given by Figure~\ref{fig:RMuller}. One can notice that the last qubit needs to be permuted to the front of the circuit. This will be done using the algorithm included online~\cite{Gitlab}. The result of this transformation is shown in Figure~\ref{fig:RMuller_opt}. After the translated circuit is obtained, it can be implemented using lattice surgery with the same steps as before. One choice for the layout of the patches is shown in Figure~\ref{fig:RM_layout} and the locations to inject $\ket{A}$ with their eventual corrective $\ket{Y}$ states are shown in Figure~\ref{fig:RM_injection}. One can see that there are many {\em empty\/} regions that exist during the preparation of the entangled state, which we anticipate can be optimized further. The total spatial requirements for this circuit are given by 60 encoded regions. The time complexity in this circuit depends on how often an erroneous $T$ has been applied.
If corrective $P$-gates have to be applied, the total time effort is higher than for the $P$-gate. Two additional steps to inject and merge corrective $\ket{Y}$-states are needed, giving a space-time volume of $1080 d^3$. This compares with a space/time volume of $1500d^3$ for the braid-based logic~\cite{FD12}.

\subsection{Bravyi-Haah code for $\ket{A}$-state distillation}

\begin{figure}
  \centering
  \includegraphics[width=0.85\columnwidth]{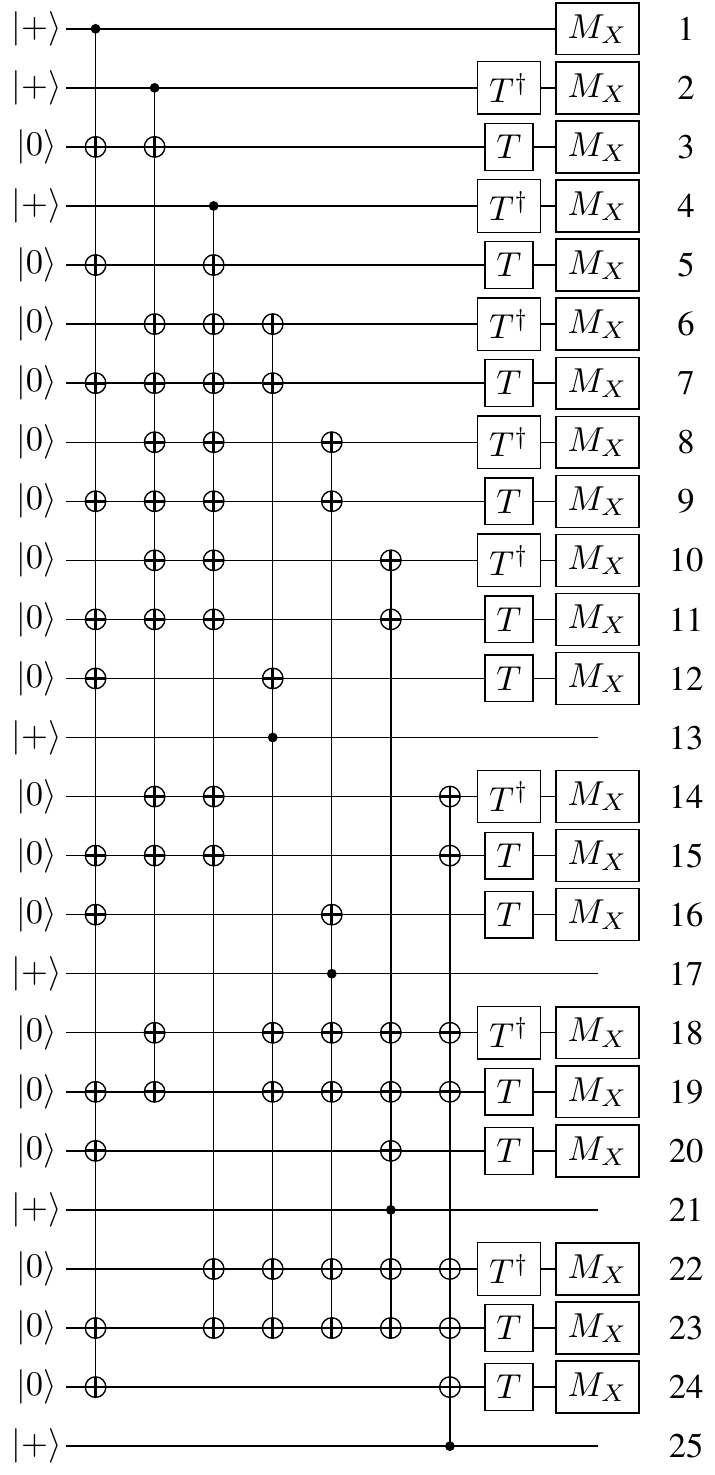}
\caption{\label{fig:BravyiHaah_circ} Circuit for the Bravyi-Haah $(3k +8)$-to-$k$ distillation code for $k = 4$. The original circuit for this code was found in Reference~\cite{bravyi_resource}. Using the algorithm given in the supplementary we translated it to this form, which can now be used for our translation to lattice surgery.}
\end{figure}

Another very promising class of distillation methods was introduced by Bravyi and Haah~\cite{Statedistill_BravyiHaah}. These are based on triorthogonal stabilizer matrices. In reference~\cite{bravyi_resource} an exemplary circuit that fulfills the triorthogonal requirement was already translated to the braiding framework. We use the same example, namely a $(3k + 8)$-to-$k$ distillation code for $\ket{A}$ with $k=4$ and compare our estimates with those for braiding.
Our translation again requires a circuit given in the inverse ICM format, whose CNOTs are then merged to as few as possible multi-target CNOTs. The result of this translation is given in Figure~\ref{fig:BravyiHaah_circ}. This circuit can now be translated to lattice surgery using a new row for each multi-target CNOT.\@ We optimized the layout of patches manually and obtained a space requirement of $7\cdot 18$ patches of surface code.
The time requirements do not change compared to the Reed-Muller code, such that this circuit needs $7d$ cycles for optimal performance, where no corrective $P$-gates are needed, and $9d$ cycles for a worst case. This calculates to a worst-case space-time volume of $2268 d^3$ in lattice surgery.
The braiding implementation of this circuit performs worse with a space-time volume of $4688 d^3$.
This is a good result; however, further research is required to obtain the scaling with $k$ for lattice surgery. In braiding, an efficient packing for arbitrary $k$ has already been found, whereas here we only performed manual optimization for this specific case.

\begin{figure}
  \centering
  \includegraphics[width=0.8\columnwidth]{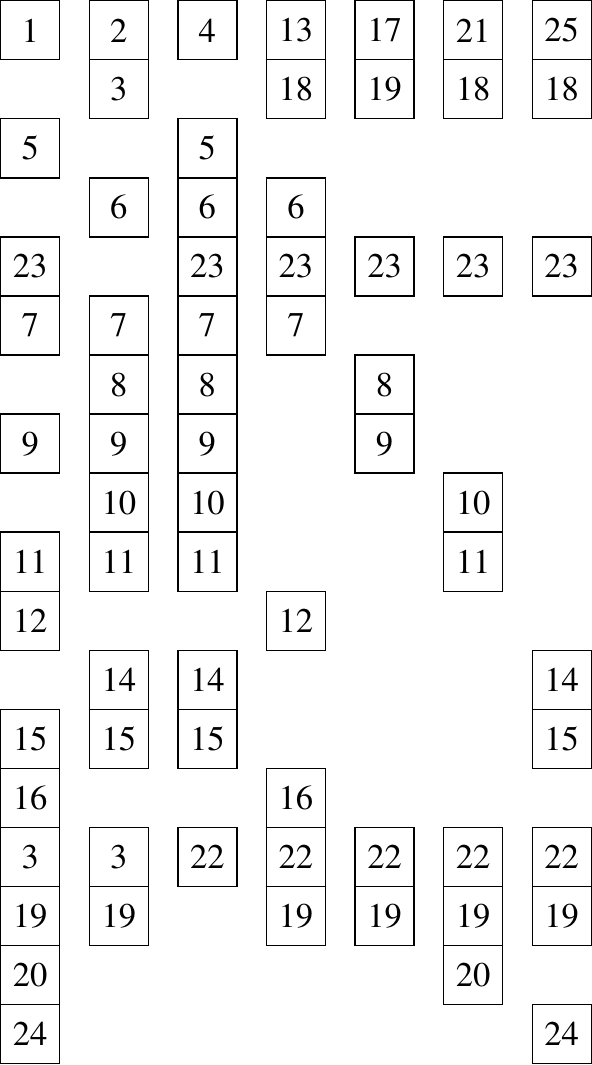}
\caption{\label{fig:Bravyi_boxrep} Continuing from the circuit in Figure~\ref{fig:BravyiHaah_circ}, we translate each multi-target CNOT into a row of surface code patches using smooth splits. The placement of these patches was again optimized manually.}
\end{figure}

\begin{figure}
  \centering
  \includegraphics[width=0.6\columnwidth]{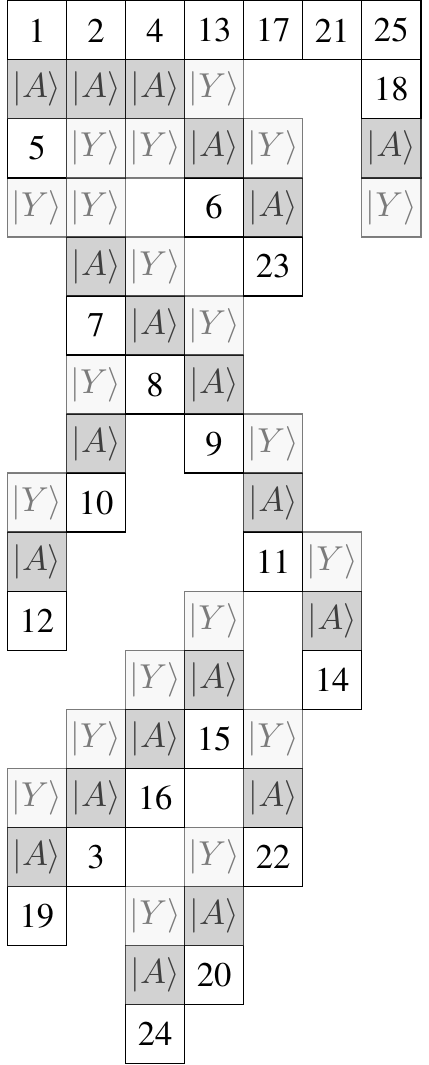}
\caption{\label{fig:Bravyi_final} After rough merges of the same numbered patches and a shrink operation we provide a placement that allows for both the application of the $T$ or $T^\dagger$ gates and their corrective $P$ gates.}
\end{figure}

\section{\label{sec:conclusion} Conclusion}
In this paper we have provided a method for compiling a fault-tolerant quantum circuit for a surface code quantum computer based on lattice surgery protocols. Using the natural operations of the lattice surgery model and a specific representation of a compatible, fault-tolerant circuit, we show via stabilizers how the Clifford part of the circuit can be directly mapped to lattice surgery protocols in the surface code.  Further work is required to optimize the arrangement and movement of encoded regions in the computer to efficiently realize any given circuit. Examples of state-distillation circuits were given, which have a comparable or better space/time volume than braid-based circuit implementations and with further optimization we expect this to decrease further. In light of recent results~\cite{Campbelldistill,Ogorman,CO16} on improved state-distillation procedures, further analysis should be conducted on how the resource cost changes with them applied to the lattice surgery model and with more efficient encodings now developed for the surface code~\cite{DIP16,NSM16}, qubit resources for an arbitrary algorithm will further decrease.

\section*{Acknowledgements}
DH is supported by the RIKEN IPA program. SJD acknowledges support from the JSPS Grant-in-aid for Challenging Exploratory Research and
and the JST ImPact project, Japan. FN was partially supported by the RIKEN iTHES Project, the MURI Center for Dynamic Magneto-Optics, the IMPACT program of JST, CREST, a Grant-in-Aid for Scientific Research (A), and a grant from the John Templeton Foundation.

\section*{Additional information}
Source code for this work can be found at \url{https://github.com/herr-d/LS_translation}
\newpage
\bibliography{biblio}
\end{document}